\documentclass[twocolumn,aps,prl,floatfix]{revtex4}
\usepackage{graphicx}
\usepackage{dcolumn}
\usepackage{url}

\begin{document}

\title{Atomic parity nonconservation in Ra$^+$}

\author{L. W. Wansbeek}
\affiliation{KVI, University of Groningen, NL-9747 AA Groningen, The Netherlands}%
\author{B. K. Sahoo}
\affiliation{KVI, University of Groningen, NL-9747 AA Groningen, The Netherlands}%
\author{R. G. E. Timmermans\footnote{Electronic address: {\tt timmermans@kvi.nl}}}
\affiliation{KVI, University of Groningen, NL-9747 AA Groningen, The Netherlands}%
\author{K. Jungmann}
\affiliation{KVI, University of Groningen, NL-9747 AA Groningen, The Netherlands}%
\author{B. P. Das}
\affiliation{Non-accelerator Particle Physics Group, Indian Institute of Astrophysics,
             Bangalore-560034, India}%
\author{D. Mukherjee}
\affiliation{Department of Physical Chemistry, Indian Association for Cultivation of Science,
             IACS, Kolkata 70032, India}
\affiliation{Raman Center for Atomic, Molecular and Optical Sciences, IACS,
             Kolkata 70032, India}
\date{\today}

\begin{abstract}
We report on a theoretical analysis of the suitability of the
$7s\,$$^2\!S_{1/2}\leftrightarrow6d\,$$^2\!D_{3/2}$ transition in singly ionized
radium to measure parity nonconservation, in the light of an experiment planned at
the KVI of the University of Groningen.
Relativistic coupled-cluster theory has been employed to perform
an {\em ab initio\/} calculation of the parity nonconserving electric dipole
amplitude of this transition, including single, double, and leading triple excitations.
We discuss the prospects for a sub-1\% precision test of the electroweak theory of
particle physics.
\end{abstract}
\maketitle

In atomic systems, parity is broken due to the exchange of the neutral vector boson
$Z^0$, which mediates the weak interaction between the atomic electrons and the quarks
in the nucleus. This atomic parity nonconservation (APNC) gives rise to small
parity nonconserving electric dipole transition amplitudes ($E1_{\text{PNC}}$).
The APNC effect gets strongly enhanced in heavy atoms
and can be measured by the interference of $E1_{\text{PNC}}$ with a
suppressed electromagnetic transition amplitude ($M1$, $E2$)~\cite{bouchiatpl74,fortsonprl93}.
The accurate measurement of the $6s\,^2\!S_{1/2}\leftrightarrow7s\,^2\!S_{1/2}$
transition in atomic $^{133}$Cs by the Boulder group~\cite{woodsc97,bennettprl99}
constitutes a precision test of the electroweak sector of the Standard Model (SM) of
particle physics~\cite{marcianoprl90}. By combining the measurement with a many-body
atomic structure calculation, the weak nuclear charge could be determined~\cite{gingespr04}.

The importance of APNC to particle physics is a strong incentive to further pursue
these challenging experiments. With the experimental and theoretical accuracies at
an impressive 0.35\% ~\cite{woodsc97,bennettprl99} and 0.5\% ~\cite{gingespr04} level,
respectively, the $^{133}$Cs result agrees with the SM prediction within one
standard deviation. Nevertheless, it is desirable to consider other candidates
for APNC studies, see {\it e.g.} Ref.~\cite{guenampla05}.
New experiments have been proposed for Cs~\cite{lintzepja07} and Fr~\cite{gomezrpp06}
atoms. Of special interest is the proposal by Fortson to measure APNC
in {\em one single\/} laser-cooled and trapped ion~\cite{fortsonprl93}. Such single-ion
experiments offer important benefits, such as long coherence times and precise
control of various systematic effects. Promising ions from the experimental and
atomic-theory point of view are heavy alkali-like ions, in particular Ba$^+$ and
Ra$^+$~\cite{koerberjpb03}.
Proof-of-principle experiments have been carried out with
$^{138}$Ba$^+$ by the Fortson group~\cite{koerberjpb03,koerberprl02,shermanprl05}.

At the TRI$\mu$P facility~\cite{bergnim06,traykovnim07} at the accelerator
institute KVI in Groningen an APNC experiment on Ra$^+$ is
in progress~\cite{trix}. An important advantage of Ra$^+$ is that all relevant
transitions are in the optical regime, {\em cf.} Fig. 1, and thus are accessible
by commercially available solid-state laser technology. The goal is to measure the
$E1_{\text{PNC}}$ amplitude of the $7s\,$$^2\!S_{1/2}\leftrightarrow 6d\,$$^2\!D_{3/2}$
transition.
We address here the question what the prospects are to push the corresponding atomic
theory below 1\%, such that the experiment can serve as a high-precision test of
the SM. We analyze various relevant properties of Ra$^+$ and assess the remaining
uncertainties.

\begin{figure}[t]
\centerline{\includegraphics*[width=0.4\textwidth]{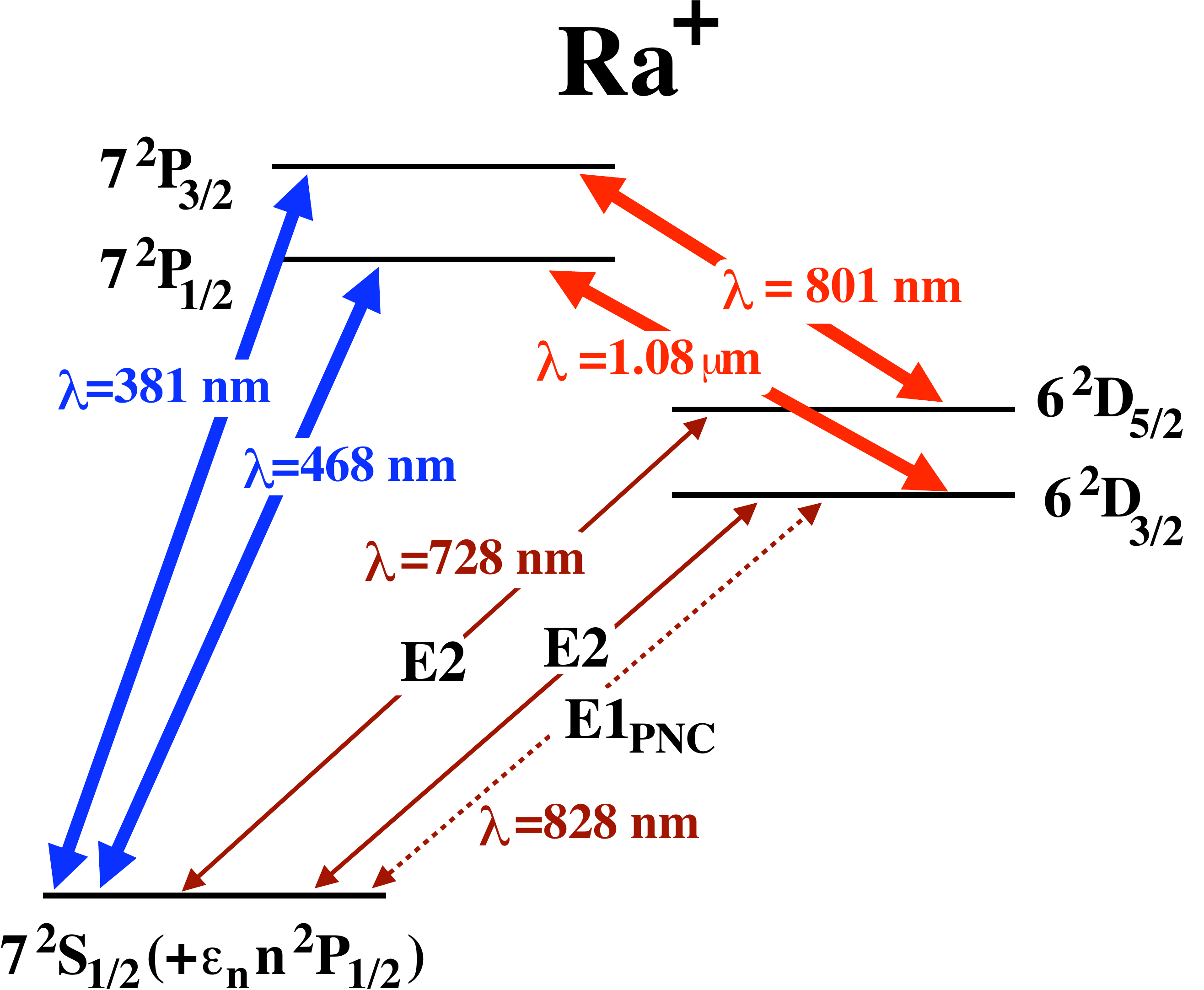}}
\caption{Relevant energy levels in Ra$^+$.}
\label{fig1}
\end{figure}
The parity-nonconserving nuclear-spin independent (NSI) interaction is due to the
electron-quark neutral weak interaction, the Hamiltonian of which is given by
\begin{equation}
   H_{\text{PNC}}^{\text{NSI}} = \frac{G_F}{2\sqrt{2}}Q_W\gamma_5\varrho_{\text{nuc}}(r) \ ,
\label{eqn1}
\end{equation}
where $G_F$ is the Fermi constant, $\varrho_{\text{nuc}}$ the nuclear density, and $\gamma_5$ is
the standard Dirac matrix; $Q_W$ is the weak nuclear charge, which is equal to
$(2Z+N)c_{1u}+(2N+Z)c_{1d}$ in terms of the coupling constants of the electron to the up and down
quarks; $Z$ and $N$ are the number of protons and neutrons. The Hamiltonian in Eq.~(\ref{eqn1})
mixes atomic states of opposite parity but with the same angular momentum. Its strength is weak
enough to consider it as a first-order perturbation. We therefore write the valence state ($v$)
atomic wave function as
\begin{equation}
   |\Psi_v\rangle = |\Psi_v^{(0)}\rangle + G_F|\Psi_v^{(1)} \rangle \ ,
\label{eqn2}
\end{equation}
where $|\Psi_v^{(0)}\rangle$ is the atomic wave function of the Dirac-Coulomb (DC)
Hamiltonian ($H_{\text{DC}}$) and $|\Psi_v^{(1)}\rangle$ is the first-order correction
due to the PNC NSI interaction.

To a first-order approximation, the $E1_{\text{PNC}}$ transition amplitude between the
$7s\,$$^2\!S_{1/2}$ ($=i$) and $6d\,$$^2\!D_{3/2}$ ($=f$) states is given by
\begin{equation}
E1_{\text{PNC}} = G_F \frac{\langle \Psi_f^{(0)}|D|\Psi_i^{(1)}\rangle
                     + \langle\Psi_f^{(1)}| D |\Psi_i^{(0)}
              \rangle}{\sqrt{\langle\Psi_f^{(0)}|\Psi_f^{(0)}
              \rangle\langle\Psi_i^{(0)}|\Psi_i^{(0)}\rangle}} \ , \label{eqn7}
\end{equation}
which, after expansion, takes the form
\begin{eqnarray}
E1_{\text{PNC}} &=& \sum_{I\ne i} \frac{\langle \Psi_f^{(0)}|D|\Psi_I^{(0)}\rangle\langle
               \Psi_I^{(0)}| H_{\text{PNC}}^{\text{NSI}} |\Psi_i^{(0)}\rangle}
       {(E_i\!-\!E_I)\sqrt{\langle\Psi_f^{(0)}|\Psi_f^{(0)}
       \rangle\langle\Psi_i^{(0)}|\Psi_i^{(0)}\rangle}} \nonumber \\
        & + &\!\!\!\!\sum_{J \ne f} \frac {\langle \Psi_f^{(0)}| H_{\text{PNC}}^{\text{NSI}}
       |\Psi_J^{(0)} \rangle \langle \Psi_J^{(0)}| D |\Psi_i^{(0)}\rangle}
       {(E_f\!-\!E_J)\sqrt{\langle\Psi_f^{(0)}|\Psi_f^{(0)}
       \rangle\langle\Psi_i^{(0)}|\Psi_i^{(0)}\rangle}} \ ,\label{eq:e1pnc}
\end{eqnarray}
where $D$ is the electric dipole ($E1$) operator, $I$ and $J$ represent the allowed
intermediate states, and $E$ is the energy of the state.
An accurate determination of $E1_{\text{PNC}}$ depends on the precision of the matrix
elements of $D$ and of $H_{\text{PNC}}^{\text{NSI}}$, and of the energy differences
between the different states. At the same time, it is also important to take all intermediate
states into account, something which is not possible in the often-used sum-over-states
approach. We therefore employ the relativistic coupled-cluster (RCC) theory, which allows
us to evaluate the properties to all orders in perturbation theory. The RCC method was
previously used to calculate APNC in $^{137}$Ba$^+$ with sub-1\% accuracy~\cite{bijayaprl06} .

We obtain the first-order wave functions of Eq.~(\ref{eqn7}) in the RCC framework
as the solution of
\begin{equation}
   G_F (H_{\text{DC}}-E_v)|\Psi_v^{(1)}\rangle =
       -H_{\text{PNC}}^{\text{NSI}}|\Psi_v^{(0)}\rangle \ ,
\label{eqn5}
\end{equation}
where $v$ stands for valence electron, which is either $i$ or $f$.
The unperturbed and perturbed wave functions are expressed as
\begin{equation}
|\Psi_v^{(0)} \rangle = \exp(T^{(0)}) \{1+S_v^{(0)}\} |\Phi_v \rangle \ ,
\end{equation}
and
\begin{eqnarray}
|\Psi_v^{(1)} \rangle &=& \exp(T') \{1+S_v'\} |\Phi_v \rangle \nonumber \\
        &=& \exp(T^{(0)}) (T^{(1)}\{ 1+ S_v^{(0)}\}+\{S_v^{(1)}\}) |\Phi_v \rangle \ ,
\end{eqnarray}
respectively, where $|\Phi_v\rangle$ is the mean-field wave function determined with the Dirac-Fock
(DF) method. $T$ and $S_v$ are the core and valence-core RCC correlation operators,
respectively, where the superscript 0 indicates in the presence of the Coulomb interaction,
the prime ($'$) indicates in the presence of both the Coulomb and APNC interaction, and 1
indicates their linear approximations. Substituting the above expressions in
Eq.~(\ref{eqn7}), we obtain
\begin{equation}
  E1_{\text{PNC}} = G_F \frac{\langle\Phi_f|C^\dagger_f \overline{D^{(0)}}C_i|\Phi_i\rangle}
                    {\sqrt{(1+N_f^{(0)})(1+N_i^{(0)})}} \ , \label{eqn8}
\end{equation}
where
\begin{eqnarray}
    N_v^{(0)} & = & \langle \Phi_v| S_v^{(0)\dagger}\exp(T^{(0)\dagger})
                    \exp(T^{(0)})S_v^{(0)} |\Phi_v\rangle \ , \\
         C_v  & = &  T^{(1)} \{1 + S_v^{(0)} \} + S_v^{(1)} \ ,
\end{eqnarray}
and $\overline{D^{(0)}} = \exp(T^{(0)\dagger})D\exp(T^{(0)})$. The matrix element
is evaluated using the generalized Wick's theorem~\cite{bijayaprl06}.

Our RCC work has two salient features. We evaluate $E1_{\text{PNC}}$ by
using the direct solution of the first-order perturbed equation as
given in Eq. (\ref{eqn5}) rather than summing over a finite number of
intermediate states \cite{blundellprd92}. The core correlation effects
modified by the parity nonconserving weak interaction are evaluated to
all orders through $T^{(1)}$ in the framework of the relativistic CCSD(T)
method. These effects cancel strongly in Cs and Fr,
where both the initial and final states are $S$-states. However, it is
essential to consider them accurately in the $S$-$D$ transitions in Ba$^+$ and Ra$^+$,
where these contributions are significant.
For our calculation, we have used numerical DF/V$^{N-1}$ orbitals to describe
the occupied and bound virtual orbitals.
The continuum states were represented by V$^{N-1}$ Gaussian-type orbitals
(GTOs)~\cite{chaudhuripra99} using the parameters $\alpha=5.25\times 10^{-3}$
and $\beta=2.73$. The finite size of the nucleus is accounted for by assuming
a Fermi charge distribution~\cite{chaudhuripra99}.
\begin{table}[b]
\setlength{\extrarowheight}{2pt}
\caption{$E1_{\text{PNC}}$ for the $7s\,^2\!S_{1/2}\leftrightarrow 6d\,^2\!D_{3/2}$
         transition in the isotope $^{226}$Ra$^+$, in units of $10^{-11}iea_0(-Q_W/N)$.}
\begin{tabular}{lcclc}
\hline\hline
\multicolumn{2}{c}{\emph{This work}} & & \multicolumn{2}{c}{Ref.~\cite{dzubapra01}} \\
    \hline
    DF&40.4  & & & \\
    CCSD& 46.1 & & Mixed-states & 42.9 \\
    CCSD(T)&46.4 & & Sum-over-states & 45.9 \\
    \hline\hline
\end{tabular}
\label{tab1}
\end{table}

In Table~\ref{tab1}, we present our RCC results for the $E1_{\text{PNC}}$ amplitude
of the $7s\,^2\!S_{1/2}\leftrightarrow 6d\,^2\!D_{3/2}$ transition in the
isotope $^{226}$Ra$^+$. Shown are the results of the DF method, of the RCC method
with single and double excitations (CCSD), and with the leading triple excitations
(CCSD(T)). The difference between the CCSD(T) and CCSD results is small.
Our best value is the CCSD(T) result $E1_{\text{PNC}}=46.4\times10^{-11}iea_0(-Q_W/N)$.
Also shown are two results of Dzuba~{\it et al.}~\cite{dzubapra01}:
in a sum-over-states approach they found $45.9\times10^{-11}iea_0(-Q_W/N)$;
in a mixed-states approach, wherein the APNC interaction explicitly mixes
states of opposite parity, they obtained $42.9\times10^{-11}iea_0(-Q_W/N)$. Neither
calculation includes structural radiation, the weak correlation potential, and
normalization of states, effects which are included by us.

The $E1_{\text{PNC}}$ amplitude for the $6s\,^2\!S_{1/2}\leftrightarrow 7s\,^2\!S_{1/2}$
transition in Cs is about 0.9$\times 10^{-11}iea_0(-Q_W/N)$~\cite{gingespr04}. Thus,
the APNC effect in Ra$^+$ is larger by a factor close to 50.
In heavy atoms, ANPC gets enhanced by the overall factor $K_rZ^2Q_W(Z,N)$,
where $Q_W\sim N\sim Z$, and $K_r$ is a relativistic factor that depends on the nuclear
charge and radius. This is the ``faster-than-$Z^3$ law''~\cite{bouchiatpl74},
which implies that Ra$^+$ is favored over Cs by a factor of about 20.
An additional factor of around 2 can be understood as follows.
For Cs (and Fr) the $S$-$S$ transition is used, for Ra$^+$ (and Ba$^+$) the $S$-$D$
transition. Since the $Z^0$-boson is very heavy, the weak interaction between electrons
and the quarks in the nucleus has (almost) zero range. The overlap of the
electrons with the nucleus is largest for the $S$ states, and thus the
mixing of the $P$-states into the $S$ states gives the major contribution to
$E1_{\text{PNC}}$. However, in Cs and Fr the initial and final $S$ states contribute
with opposite signs, which leads to a significant cancellation in $E1_{\text{PNC}}$.
In fact, for Cs there are three dominant terms in the sum over the states, which add up
to a total value that is half the size of the largest individual term~\cite{gingespr04}.

The $S$-$D$ transitions in Ba$^+$ and Ra$^+$ do not suffer from such a cancellation,
since the contribution from the $D$-state to APNC is small.
In Table~\ref{tab2}, we analyze which intermediate states contribute most to
the total sum. Clearly, in contrast to the Cs $S$-$S$ case, the sum is strongly
dominated by one term: the contribution from the $7p\,^2\!P_{1/2}$ state. These
qualitative results are robust, they are consistent with the findings of
Ref.~\cite{dzubapra01}, and they are, in fact, already borne out by a simple
calculation with quantum-defect theory, analogous to Ref.~\cite{bouchiatjp7475}.
This simple estimate gives for Cs, Fr, Ba$^+$, and Ra$^+$ results accurate to some 10\%;
for Ra$^+$ we find $E1_{\text{PNC}}=45(4)\times10^{-11}iea_0(-Q_W/N)$.

\begin{table}[b]
\setlength{\extrarowheight}{2pt}
\caption{The contributions to the $E1_{\text{PNC}}$ from the different $P$-states (\%).}
\begin{ruledtabular}
\begin{tabular}{lcrlcr}
 State  & type & \% & State  & type & \% \\
  \hline
   $6p\,^2\!P_{1/2}$ &  core &    8.7 &  $8p\,^2\!P_{1/2}$ &   bound   & $-$3.3 \\
   $6p\,^2\!P_{3/2}$ &  core &  $-$15 &  $9p\,^2\!P_{1/2}$ &   bound   & $-$0.7 \\
   $7p\,^2\!P_{1/2}$ & bound &    111 & $10p\,^2\!P_{1/2}$ & continuum & $-$0.1 \\
   $7p\,^2\!P_{3/2}$ & bound & $-$2.6 & $11p\,^2\!P_{1/2}$ & continuum &    1.1 \\
\end{tabular}
\end{ruledtabular}
\label{tab2}
\end{table}

In Table~\ref{tab3}, we present our results for the excitation energies, $E1$ transition
amplitudes, and hyperfine constants for the relevant transitions and states in Ra$^+$. We also list
experimental values where available. For the excitation energies, we compare to the only available
spectroscopy measurement~\cite{rasmussenzp33}, which dates back to 1933. For the $E1$ transition
amplitudes, for which there are no experimental data, we list the results of Ref.~\cite{dzubapra01}
and of our previous work~\cite{sahoopra07} using GTOs. Therein, the lifetimes of the metastable
$D$-states were calculated to be 0.627(4) s for $6d\,^2\!D_{3/2}$ and 0.297(4) s for
$6d\,^2\!D_{5/2}$.

\begin{table}[t]
\caption{Excitation energies, $E1$ transition amplitudes, and $A_I/g_I$ for
different low-lying states of Ra$^+$.}
\begin{ruledtabular}
\begin{tabular}{lllll}
& $7s\,^2\!S_{1/2}$ & $7s\,^2\!S_{1/2}$ & $6d\,^2\!D_{3/2}$ & $6d\,^2\!D_{3/2}$ \\
\raisebox{1ex}[0pt]{Transition}
& $7p\,^2\!P_{1/2}$ & $7p\,^2\!P_{3/2}$ & $7p\,^2\!P_{1/2}$ & $7p\,^2\!P_{3/2}$ \\
\hline\tabularnewline
\multicolumn{2}{l}{\emph{Excitation energy} [cm$^{-1}$]} &  &  & \\
This work                       & 21509 & 26440 & 9734 & 14665 \\
Experiment~\cite{rasmussenzp33} & 21351 & 26209 & 9267 & 14125 \\
\tabularnewline
\multicolumn{2}{l}{$E1$ \emph{transition amplitude} [a.u.]}   &  &  & \\
This work              & 3.31  & 4.58  & 3.68  & 1.56  \\
Ref.~\cite{dzubapra01} & 3.223 & 4.477 & 3.363 & 1.504 \\
GTOs~\cite{sahoopra07} & 3.28  & 4.54  & 3.64  & 1.54  \\
\tabularnewline
State & $7s\,^2\!S_{1/2}$ & $7p\,^2\!P_{1/2}$ & $7p\,^2\!P_{3/2}$ & $6d\,^2\!D_{3/2}$ \\
\hline \tabularnewline
\multicolumn{4}{l}{ \emph{Hyperfine interaction constant} ($A_I/g_I$) [MHz]} & \\
This work               & 19689.37 & 3713.75 & 312.91 & 441.67 \\
Experiment \cite{neu89} & 18772    & 3691    & 314.12 &  -     \\
\end{tabular}
\end{ruledtabular}
\label{tab3}
\end{table}

Since the hyperfine structure is a good probe of the wave functions at the nucleus,
we have, in order to estimate the accuracy of the $H_{\text{PNC}}^{\text{NSI}}$ matrix
elements, calculated the ratio between the magnetic dipole hyperfine structure ($A$)
and the nuclear gyromagnetic ($g$) constants, neglecting isotope effects, and compared
these with experimental results for Ra$^+$ from ISOLDE~\cite{neu89,wendt87}.
Our calculated value for $[A_I/g_I(7S_{1/2})A_I/g_I(7P_{1/2})]^{1/2}$ differs by
3\% from the experimental value, which is a reasonable estimate for the dominant
uncertainty in the atomic theory. Thus, our best value for the parity nonconserving
$E1$ amplitude in Ra$^+$ is
$E1_{\text{PNC}}=46.4(1.4)\times10^{-11}iea_0(-Q_W/N)$.

It appears feasible to push the accuracy of the atomic theory for Ra$^+$ to the sub-1\% level.
Improvements along several lines are in progress. The Breit
interaction~\cite{dereviankoprl00,dzubapra06} and QED corrections, which contribute around 1\%,
need to be included. The neutron-skin effect~\cite{brownA08}, which also contributes at the sub-1\%
level, has to be investigated. However, at the same time it is clear that experimental information
to test the atomic theory is severely lacking. Not all relevant energy levels are
known~\cite{rasmussenzp33}, there is no experimental information on the $E1$ transition strengths,
nor on the lifetimes of the $D$-states. It is highly desirable to have more experimental data on
the magnetic dipole and electric quadrupole hyperfine constants $A$ and $B$ for the various Ra$^+$
isotopes. The extraction of these constants~\cite{neu89,wendt87} is model dependent, and ideally
one would like to use a single consistent {\it ab initio\/} framework for this.

At the TRI$\mu$P facility, radium isotopes can be produced in
fusion and evaporation or spallation reactions. The ions can be collected
in a radio-frequency trap where they can be laser-cooled on the
$7s\,^2\!S_{1/2}\leftrightarrow7p\,^2\!P_{1/2}$ resonance line at 468 nm, with
repumping via the $6d\,^2\!D_{3/2}\leftrightarrow7p\,^2\!P_{1/2}$ transition at
1.08 $\mu$m, for which strong lasers are available. They will then be transferred
to a miniature trap for the single-ion experiment, where techniques similar to
Refs.~\cite{koerberjpb03,koerberprl02,shermanprl05} will be applied to perform
the measurements. In particular, the $7s\,^2\!S_{1/2}\leftrightarrow7p\,^2\!P_{3/2}$
and $6d\,^2\!D_{5/2}\leftrightarrow7p\,^2\!P_{3/2}$ transitions at 381 nm and 801 nm,
respectively, can be used for ``shelving''~\cite{koerberjpb03}.

\begin{table}[t]
\setlength{\extrarowheight}{2pt}
\caption{The properties of the isotopes of Ra$^+$ suitable for a single-ion APNC experiment.
         $A$ is the mass number, $I$ is the nuclear spin, and $\tau_{1/2}$ the half-life
         time.}
\begin{tabular}{ccll}
\hline\hline
 $A$   &$I$&$\tau_{1/2}$&Possible production reaction \\ \hline
213&1/2$^-$&2.74(6) min&$^{208}$Pb + $^{12}$C $\rightarrow$ $^{213}$Ra + 7n \\
223&3/2$^+$&11.43(5) d&p + $^{232}$Th $\rightarrow$ $^{223}$Ra + $\!^A$X + $\!a$n + $\!b$p \\
224&0$^+$&3.6319(23) d&p + $^{232}$Th $\rightarrow$ $^{224}$Ra + $\!^A$X + $\!a$n + $\!b$p \\
225&1/2$^+$&14.9(2) d&$^{229}$Th $\rightarrow$ $^{225}$Ra + $\alpha$ \\
226&0$^+$&1600(7) y&Commercially available \\
227&3/2$^+$&42.2(5) min&p + $^{232}$Th $\rightarrow$ $^{227}$Ra + $\!^A$X + $\!a$n + $\!b$p \\
\hline\hline
\end{tabular}
\label{tab4}
\end{table}
A list of the Ra$^+$ isotopes suitable for a single-ion experiment is shown in Table~\ref{tab4}.
A half-life of the order of seconds is required for a high-precision single-ion experiment,
but, on the other hand, it should not exceed a few days, so as to avoid long-lived radioactive
contamination of the core equipment. Good candidates therefore are the odd
isotopes $^{213}$Ra$^+$ and $^{227}$Ra$^+$, and the even isotope $^{224}$Ra$^+$. (Since the odd
isotopes have a nonzero nuclear spin, the nuclear spin-dependent weak interaction will
contribute to $E1_{\text{PNC}}$~\cite{geethapra98}). The isotopes listed can
all be produced at TRI$\mu$P, and the experimental data required to constrain the atomic
theory can be measured there with laser spectroscopy. Since multiple Ra$^+$ isotopes will be
available, the possibility exists to measure APNC in a chain of isotopes, which can
help to eliminate remaining uncertainties in the atomic theory~\cite{dzubazpd86}.

In conclusion, Ra$^+$ appears to be an excellent candidate for an APNC experiment,
since $E1_{\text{PNC}}$ is large, the required lasers are all at convenient wavelengths,
and one can exploit the high-precision techniques of single-ion trapping. The atomic theory
needed for the interpretation of the experiment could reach an accuracy better than 1\%,
but precise experimental data for the relevant atomic properties are mandatory to achieve
such a benchmark. The prospects for APNC in Ra$^+$ as a precise test of the SM
look promising.

Part of this work was supported by the Dutch Stichting voor Fundamenteel Onderzoek der
Materie (FOM) under program 48 (TRI$\mu$P) and project 06PR2499. The calculations were
carried out using the Tera-flop Supercomputer in C-DAC, Bangalore.

\bibliography{cc2}
\end{document}